\documentstyle[aps,preprint]{revtex}
\begin{document}
\draft
\title{Anisotropic Scaling in Depinning of a Flux Line}
\author{Deniz Erta\c s and Mehran Kardar}
\address{Department of Physics,
Massachusetts Institute of Technology,
Cambridge, Massachusetts 02139}
\date{\today}
\maketitle
\begin{abstract}
We study the depinning of a flux line by analytical and numerical
methods applied to a phenomenological equation of motion.
Transverse fluctuations  do not influence the critical behavior of  
the longitudinal component, justifying ``planar approximations".
In an isotropic medium, longitudinal fluctuations have
a roughness exponent $\zeta_\parallel=1$, and relax
with a dynamic exponent $z_\parallel\approx4/3$; transverse  
fluctuations
are suppressed ($\zeta_\perp=1/2<\zeta_\parallel$), and relax
more slowly, with $z_\perp=z_\parallel+1$.
Anisotropy in the depinning threshold, or orientational dependence
of force-force correlations, lead to new  universality classes.
\end{abstract}

The pinning of flux lines (FLs) in Type-II superconductors is of
fundamental importance to many technological applications that
require large critical currents\cite{review}.
Upon application of an external current density $J$, the motion of
FLs due to the Lorentz force causes undesirable dissipation of
supercurrents. Major increases in the critical current
density $J_c$ of a sample are achieved when the FLs are pinned
to impurities. There are many recent studies,  both
experimental\cite{VGexp,colexp}
and theoretical\cite{VGtheo,coltheo}, on collective pinning of FL's  
to
point or columnar defects. Another consequence of impurities is
the strongly nonlinear behavior of the current slightly above the
depinning threshold, as the FLs start to move across the sample.
Recent numerical simulations have concentrated on the low temperature
behavior of a single FL near depinning\cite{Enomoto,Dong,Tang},
mostly ignoring fluctuations transverse
to the plane defined by the magnetic field and the Lorentz force.
Common signatures of the depinning transition from $J<J_c$ to
$J>J_c$ include a broad band ($f^{-a}$ type) voltage noise spectrum,
and self-similar fluctuations of the FL profile.

The FL provides yet another example of a broad class of systems that
exhibit a depinning transition; a dynamical critical
phenomenon seen in Charge-Density Waves (CDWs)
\cite{Fisher85,Robbins,NF92,Middleton},
interfaces in random media\cite{NSTL,NF93},
and contact lines\cite{Joanny,EKcl}.
There has recently been much theoretical progress in calculating the  
critical
exponents of such transitions by renormalization-group (RG) methods.
In this Letter, we extend these methods to study the full  
three-dimensional
dynamics of a single FL at low temperatures.
We show that fluctuations transverse to the average
motion of the FL do not change the scaling properties of the
longitudinal component. In turn, the transverse fluctuations
are suppressed, and relax much more slowly (with a larger dynamic  
exponent).
Thus, we not only formally justify ignoring transverse fluctuations
(i.e. planar approximations) in numerical simulations near the
depinning transition, but also demonstrate the rich anisotropic
scaling properties of this system.

In the simplest case of an isotropic pinning potential, the  
longitudinal
fluctuations have roughness exponent $\zeta_\parallel=1$, and   
dynamic
exponent $z_\parallel\approx4/3$, while  
$\zeta_\perp=1/2<\zeta_\parallel$,
and  $z_\perp=z_\parallel+1$ for the transverse components.
When the pinning potential is anisotropic and the FL is driven along  
a
non-symmetric direction, transverse fluctuations are further  
suppressed
to $\zeta_\perp=0$, possibly with logarithmic deviations.
If force-force correlations depend on the orientation of the FL as  
well as
its position, the critical behavior falls into yet another  
universality
class, possibly controlled by directed percolation clusters. Thus,  
the
depinning
behavior is quite rich, described by a number of distinct  
universality classes.

We describe the shape of the FL at a given time $t$ by ${\bf r}(x,t)$,
where $x$ is along the magnetic field $B$, and the unit vector
${\bf e}_\parallel$ is along the Lorentz force ${\bf F}$.
(See Fig. \ref{geometry}.) Point impurities are modeled by a random
potential $V(x,{\bf r})$, with zero mean and short-range
correlations. In the presence of impurities and a bulk Lorentz
force ${\bf F}$, the energy of a FL with small fluctuations is,

\begin{equation}
\label{Hamiltonian}
{\cal H} = \int dx\left\{\frac{1}{2}(\partial_x {\bf r})^2
+V\left(x,{\bf r}(x,t)\right)-{\bf r}(x,t)\cdot{\bf F}\right\}.
\end{equation}
The simplest possible Langevin equation for the FL, consistent with
{\it local, dissipative dynamics}, is
\begin{equation}
\label{motion}
\mu^{-1}\frac{\partial {\bf r}}{\partial t} = -\frac{\delta {\cal  
H}}{\delta {\bf r}}=
\partial_x^2 {\bf r}+{\bf f} \left(x,{\bf r}(x,t)\right)+{\bf F},
\end{equation}
where $\mu$ is the mobility of the FL, and ${\bf f}=-\nabla_{\bf r} V$.
The potential $V(x,{\bf r})$ need not be
isotropic. For example, in a single crystal of ceramic  
superconductors
with the field along the oxide planes, it will be easier to move
the FL along the planes. This leads to a pinning threshold that
depends on the orientation of the force. Anisotropy also modifies the  
line
tension, and the elastic term in Eq.~(\ref{motion}) is in general  
multiplied by
a non-diagonal matrix $K_{\alpha\beta}$.
The random force ${\bf f}(x,{\bf r})$, can be taken to have zero
mean with correlations
\begin{equation}
\label{statistics}
\langle f_\alpha(x,{\bf r})f_\gamma(x',{\bf r}')\rangle=
\delta(x-x')\Delta_{\alpha\gamma}({\bf r}-{\bf r}').
\end{equation}
We shall focus mostly on the isotropic case,
with $\Delta_{\alpha\gamma}({\bf r}-{\bf r}')=
\delta_{\alpha\gamma}\Delta(|{\bf r}-{\bf r}'|)$,
where $\Delta$ is a function that decays rapidly for large
values of its argument.

While the flux line is pinned by impurities when $F<F_c$,
for $F$ slightly above threshold, we expect the average velocity
$v=|{\bf v}|$ to scale as
\begin{equation}
\label{velocity}
v\sim(F-F_c)^\beta,
\end{equation}
where $\beta$ is the velocity exponent. Superposed on the  
steady
advance of the FL are rapid  ``jumps" as portions of the line depin
from strong pinning centers. Such jumps are similar to avalanches
in other slowly driven systems, and have a power-law distribution
in size, cut off at a  characteristic correlation length $\xi$.
On approaching the threshold, $\xi$ diverges as
\begin{equation}
\xi\sim (F-F_c)^{-\nu},
\end{equation}
defining a correlation length exponent $\nu$. At length scales
up to $\xi$, the correlated fluctuations satisfy the dynamic scaling  
form,
\begin{mathletters}
\label{scaling}
\begin{eqnarray}
\langle[r_\parallel(x,t)-r_\parallel(0,0)]^2\rangle
&=& |x|^{2\zeta_\parallel}g_\parallel(t/|x|^{z_\parallel}),\\
\langle[r_\perp(x,t)-r_\perp(0,0)]^2\rangle
&=& |x|^{2\zeta_\perp}g_\perp(t/|x|^{z_\perp}),
\end{eqnarray}
\end{mathletters}
where $\zeta_\alpha$ and $z_\alpha$ are the roughness and
dynamic exponents, respectively. The scaling functions
$g_\alpha$ go to a constant as their arguments approach 0.
Beyond the length scale $\xi$, different regions of the FL
depin more or less independently and  the system crosses over
to a moving state, described by different exponents $\zeta_+$,and  
$z_+$.

The major difference of this model from previously studied ones
is that the position of the flux line, ${\bf r}(x,t)$, is now a
2-dimensional vector instead of a scalar; fluctuating along both
${\bf e}_\parallel$ and ${\bf e}_\perp$ directions.
One consequence is that a ``no passing" rule\cite{Middleton},  
applicable to CDWs and interfaces, does not apply to FLs. 
It is possible to have coexistence of moving and stationary FLs 
in particular realizations of the random potential.

The effects of transverse fluctuations $r_\perp$ for large driving  
forces, when the impurities act as white noise, were studied  
earlier\cite{EKlines}, indicating a rich dynamical phase diagram. 
How do these transverse fluctuations scale near the depinning 
transition, and do they in turn influence the critical dynamics of 
longitudinal fluctuations near threshold?
The answer to the second question can be obtained by the following  
qualitative argument:
Consider Eq.~(\ref{motion}) for a particular realization of randomness
${\bf f}(x,{\bf r})$. Assuming that portions of the FL always
move in the forward direction\cite{backwards}, there is a unique
point $r_\perp(x,r_\parallel)$ that is visited by the line
for given coordinates $(x,r_\parallel)$. We construct a new force
field $f'$ on a two dimensional space $(x,r_\parallel)$
through $f'(x,r_\parallel)\equiv f_\parallel\left(x,r_\parallel,
r_\perp(x,r_\parallel)\right)$. It is then clear that the dynamics
of the longitudinal component $r_\parallel(x,t)$ in a given force field
${\bf f}(x,{\bf r})$ is identical to the dynamics
of $r_\parallel(x,t)$ in a force field $f'(x,r_\parallel)$, with
$r_\perp$ set to zero. It is quite plausible that, after
averaging over all ${\bf f}$, the correlations in $f'$ will
also be short-ranged, albeit different from those of
${\bf f}$. Thus, the scaling of longitudinal fluctuations of the
depinning FL will not change upon taking into account transverse
fluctuations. However, the question of how these transverse
fluctuations scale still remains.

Certain statistical symmetries of the system restrict the form of
response and correlation functions. For example, Eq.~(\ref{motion})
has statistical space- and time-translational invariance, which  
enables
us to work in Fourier space, i.e. $(x,t)\to(q,\omega)$.
For an {\it isotropic} medium, ${\bf F}$ and ${\bf v}$ are parallel to
each other, i.e., ${\bf v}({\bf F})=v(F){\bf \hat F}$, where ${\bf \hat F}$  
is
the unit vector along {\bf F}. Furthermore, all
expectation values involving odd powers of a transverse
component are identically zero due to the statistical invariance  
under the
transformation $r_\perp\to -r_\perp$. Thus, linear
response and two-point correlation functions  are {\it diagonal}.
The introduced critical exponents are then related through scaling  
identities.
These can be derived from the
linear response to an infinitesimal external force field
${\mbox{\boldmath$\varepsilon$}}(q,\omega)$,
\begin{equation}
\label{chi}
\chi_{\alpha\beta}(q,\omega)=\left<\frac{\partial r_\alpha(q,\omega)}
{\partial\varepsilon_\beta(q,\omega)}\right>\equiv  
\delta_{\alpha\beta}\chi_\alpha(q,\omega),
\end{equation}
in the $(q,\omega)\to(0,0)$ limit. Eq.(\ref{motion})
is statistically invariant under the transformation ${\bf F}\to{\bf F}+
{\mbox{\boldmath$\varepsilon$}}(q),  
{\bf r}(q,\omega)\to{\bf r}(q,\omega)
+q^{-2}{\mbox{\boldmath$\varepsilon$}}(q)$.
Thus, the static linear response has the form
$\chi_\parallel(q,\omega=0)=\chi_\perp(q,\omega=0)=q^{-2}$.
Since $\varepsilon_\parallel$ scales like the applied force,
the form of the linear response at the correlation length $\xi$
gives the exponent identity
\begin{equation}\
\label{nuexp}
\zeta_\parallel+1/\nu=2.
\end{equation}

Considering the transverse linear response
seems to imply $\zeta_\perp=\zeta_\parallel$. However,
the static part of the transverse linear response is
irrelevant at the critical RG fixed point, since
$z_\perp>z_\parallel$, as shown below.
When a slowly varying uniform external force 
${\mbox{\boldmath$\varepsilon$}}(t)$ is applied,
the FL responds as if the instantaneus external force
${\bf F}+{\mbox{\boldmath$\varepsilon$}}$ is a constant, 
acquring an average velocity,
\begin{equation}
\langle\partial_t r_\alpha\rangle=v_\alpha({\bf F}
+{\mbox{\boldmath$\varepsilon$}})\approx
v_\alpha({\bf F})+\frac{\partial v_\alpha}{\partial F_\gamma}
\varepsilon_\gamma.
\end{equation}
Substituting $\partial v_\parallel/\partial F_\parallel=dv/dF$ and
$\partial v_\perp/\partial F_\perp=v/F$, and Fourier transforming,  
gives
\begin{eqnarray}
\label{chipara}
\chi_\parallel(q=0,\omega) &=& \frac{1}{-i\omega(dv/dF)^{-1}
+O(\omega^2)}, \\
\label{chiperp}
\chi_\perp(q=0,\omega) &=& \frac{1}{-i\omega(v/F)^{-1}+O(\omega^2)}.
\end{eqnarray}
Combining these with the static response, we see that the  
characteristic
relaxation times of fluctuations with wavelength $\xi$ are
\begin{eqnarray}
\tau_\parallel(q=\xi^{-1}) &\sim& \left(q^2\frac{dv}{dF}\right)^{-1}
\sim\xi^{2+(\beta-1)/\nu}\sim\xi^{z_\parallel}, \\
\tau_\perp(q=\xi^{-1}) &\sim& \left(q^2\frac{v}{F}\right)^{-1}
\sim\xi^{2+\beta/\nu}\sim\xi^{z_\perp},
\end{eqnarray}
which, using Eq.(\ref{nuexp}), yield the scaling relations
\begin{eqnarray}
\label{betaexp}
\beta=(z_\parallel-\zeta_\parallel)\nu, \\
\label{zperpexp}
z_\perp=z_\parallel+1/\nu.
\end{eqnarray}
We already see that the dynamic relaxation of transverse
fluctuations is much slower than longitudinal ones.
All critical exponents can be calculated from $\zeta_\parallel$,
$\zeta_\perp$, and $z_\parallel$, by using Eqs.(\ref{nuexp}),
(\ref{betaexp}), and (\ref{zperpexp}).

Equation (\ref{motion}) can be analyzed using the formalism of
Martin, Siggia, and Rose (MSR)\cite{MSR}. Ignoring transverse
fluctuations, and generalizing to  $d$ dimensional internal  
coordinates
${\bf x}\in\Re^d$, leads to an interface
depinning model which was studied by Nattermann, Stepanow, Tang, and
Leschhorn (NSTL)\cite{NSTL}, and by Narayan and Fisher
(NF)\cite{NF93}. The RG treatment indicates that impurity
disorder becomes relevant for dimensions $d\leq4$, and the
critical exponents in $d=4-\epsilon$ dimensions are given to one-loop
order as $\zeta=\epsilon/3$, $z=2-2\epsilon/9$.
NSTL obtained this result by directly averaging the MSR
generating functional $Z$, and calculating the renormalization
of the force-force correlation function $\Delta(r)$, perturbatively
around the freely moving interface $[\Delta(r)=0]$.
NF, on the other hand, used a perturbative expansion of $Z$, around a
saddle point corresponding to a mean-field approximation to
Eq.(\ref{motion})\cite{SZ}, which involved {\it temporal}
force-force correlations $C(vt)$. They argue that a
conventional low-frequency analysis is not sufficient to determine
critical exponents. They  also suggest that the roughness
exponent is equal to $\epsilon/3$ to all orders in perturbation
theory.

Following the approach of NF, we employ a perturbative expansion
of the disorder-averaged MSR partition function around a mean-field
solution for cusped impurity potentials\cite{NF93}. All terms in the
expansion involving longitudinal fluctuations are identical to the
interface case, thus we obtain the same critical exponents for
longitudinal fluctuations, i.e., $\zeta_\parallel=\epsilon/3$,
$z_\parallel=2-2\epsilon/9+O(\epsilon^2)$. Furthermore, {\it for  
isotropic
potentials}, the renormalization of transverse temporal force-force
correlations $C_\perp(vt)$ yields a transverse roughness exponent
$\zeta_\perp=5\zeta_\parallel/2-2$, to all orders in perturbation
theory. The details of this calculation will be given elsewhere.
For the FL $(\epsilon=3)$, the critical exponents are then given by
\begin{equation}
\begin{array}{lll}
\zeta_\parallel=1,&z_\parallel\approx4/3,&\nu=1, \\
\beta\approx1/3,&\zeta_\perp=1/2,&z_\perp\approx7/3.
\end{array}
\end{equation}

To test the scaling forms and  exponents predicted by  
Eqs.(\ref{velocity})
and (\ref{scaling}), we numerically integrated Eq.(\ref{motion}),
discretized in coordinates $x$ and $t$.
Free boundary conditions were used for system sizes of up to
2048, with a grid spacing $\Delta x=1$ and a time step $\Delta  
t=0.02$.
Time averages were evaluated {\it after}
the system reached steady state. Periodic boundary conditions gave
similar results, but with larger finite size effects. Smaller grid  
sizes
did not change the results considerably.
The behavior of $v(F)$  seems to fit the scaling form of
Eq.(\ref{velocity}) with an exponent $\beta\approx0.3$, but is also
consistent with a logarithmic dependence on the reduced force, i.e.,
$\beta=0$. The same behavior was observed by Dong {\em et al.}
in a recent simulation of the $1+1$ dimensional geometry\cite{Dong}.
Since $z_\parallel$, and consequently $\beta$, is known only to
first order in $\epsilon$, higher order corrections are expected.
By looking at equal time correlation functions 
(see Fig. \ref{xcorr}) , we  
find that
transverse
fluctuations are strongly suppressed, and that the roughness
exponents are equal to our theoretical estimates within statistical
accuracy. The excellent agreement for
$\epsilon=3$ suggests that the theoretical estimates are indeed  
exact.

The anisotropy in critical exponents may be observed in a rectangular
Hall geometry, by measuring the noise power spectra  
$S_\parallel(\omega)$
and $S_\perp(\omega)$ of normal and Hall voltages, respectively.
In a conventional type-II superconductor with point defects, at low
temperatures and near depinning, Eqs.(\ref{scaling}) suggest 
that\cite{Tang}
$S_\alpha(\omega)\sim\omega^{-a_\alpha}$, where $a_\alpha=(2
\zeta_\alpha+1)/z_\alpha-1$. Thus,  
$S_\parallel(\omega)\sim\omega^{-5/4}$,
whereas $S_\perp(\omega)\sim\omega^{1/7}$ at small $\omega$.

The potential pinning the FL in a single superconducting crystal is
likely to be highly {\it anisotropic}. For example, consider a  
magnetic
field parallel to the copper oxide planes of a ceramic  
superconductor.
The threshold force then depends on its orientation,
with depinning easiest along the copper oxide planes.
Eqs.(\ref{chi}), (\ref{chipara}) and (\ref{chiperp}) have to be 
modified, since ${\bf v}$ and ${\bf F}$ are no
longer parallel (except along the axes with
${\bf r}\to -{\bf r}$ symmetry), and the linear response function is not
diagonal. The RG analysis is more cumbersome:
We find that, for depinning along a non-symmetric direction,
the longitudinal exponents are not modified (in agreement
with the argument presented earlier), while the transverse
fluctuations are further suppressed to $\zeta_\perp=
2\zeta_\parallel-2$ (equal to zero for  
$\zeta_\parallel=1$)\cite{footnote}.
Relaxation of transverse modes are still characterized by
$z_\perp=z_\parallel+1/\nu$, and the exponent identity (\ref{nuexp})
also holds. Surprisingly, the exponents for depinning along
axes of reflection symmetry are the same as the isotropic case.

Another feature of the problem is the possible dependence of
the force-force correlations $\Delta$ on the orientation $\partial_x  
{\bf r}$
of the FL in anisotropic superconductors. As in the case of
interfaces\cite{TK}, such dependence leads to introduction of
additional relevant terms in the MSR partition function, and  
invalidates
the arguments leading to  
Eqs.(\ref{nuexp},\ref{betaexp},\ref{zperpexp}).
The analogy to interfaces
suggests that the longitudinal exponents for $d=1$ are controlled by
directed percolation clusters\cite{Sneppen}, with 
$\zeta_\parallel\approx 0.63$. Since no
perturbative fixed point is present in this case, it is not
clear how to explore the behavior of transverse fluctuations
systematically.

In conclusion, we studied the dynamical critical behavior of a
single depinning FL in a type-II superconductor at low temperatures.
Using powerful symmetry arguments, we demonstrated the anisotropy
in both the configurational and relaxational properties of the FL,
which was confirmed by a formal RG treatment and
numerical simulations. This justifies the ``planar" approximation,
widely used in numerical simulations of a depinning FL. Due to  
possible
anisotropies in the pinning force, the depinning behavior is quite
rich, encompassing a number of different universality classes.
In the fully isotropic case, RG calculations
suggest anisotropic scaling with $z_\perp=z_\parallel+1$, and
$\zeta_\perp=1/2$; the latter is confirmed by numerical simulations.
Anisotropic potentials, and orientation dependent force correlations
lead to a new exponents, some of which we have obtained.
As the current density is further increased, other nonlinear terms
become relevant and change the scaling of fluctuations. The effects
of such terms were considered previously in Ref.\cite{EKlines} for a
single FL, and in Ref.\cite{Hwa} for a dense collection of FLs. The
critical behavior of an interacting ensemble of FLs near depinning
is likely to be further influenced by complicated entanglement  
effects.

We have benefited from discussions with O.~Narayan and L.-H.~Tang.
This research was
supported by grants from the NSF (DMR-93-03667 and PYI/DMR-89-58061),
and the MIT/INTEVEP colloborative program.

\begin{figure}
\caption{Geometry of the fluctuating Flux Line.}
\label{geometry}
\end{figure}

\begin{figure}
\caption{Equal time correlations as  functions of  separation $x$,
for a system of size 2048 at $(F-F_c)/F_c\approx0.01$. The observed  
roughness
exponents are close to the theoretical predictions of
$\zeta_\parallel=1,\;\zeta_\perp=0.5$,  shown by solid lines
for comparison.}
\label{xcorr}
\end{figure}


\begin{references}
\bibitem{backwards}This is not strictly true, but backward motion  
happens very rarely, and only at short length scales.
\end{references}
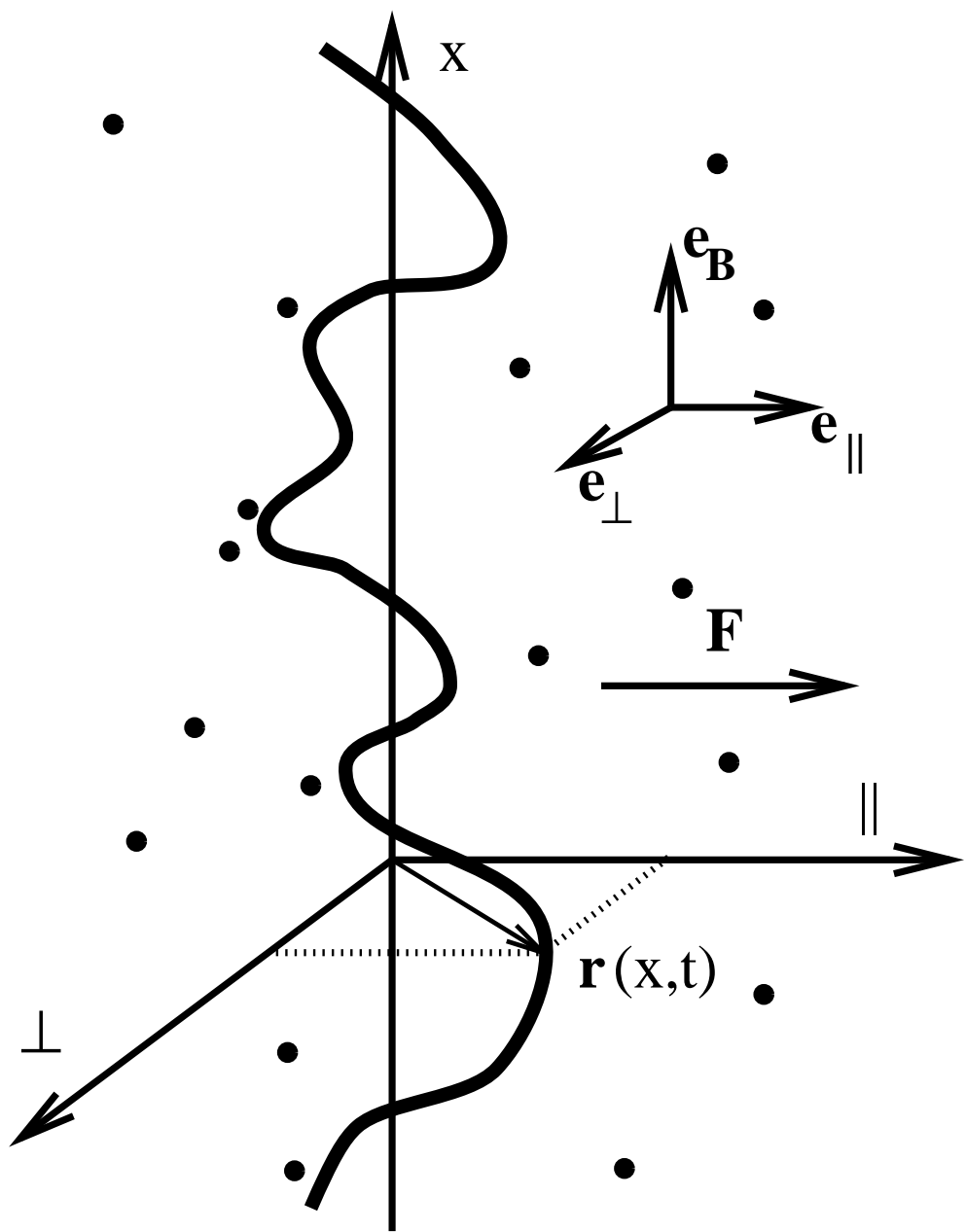
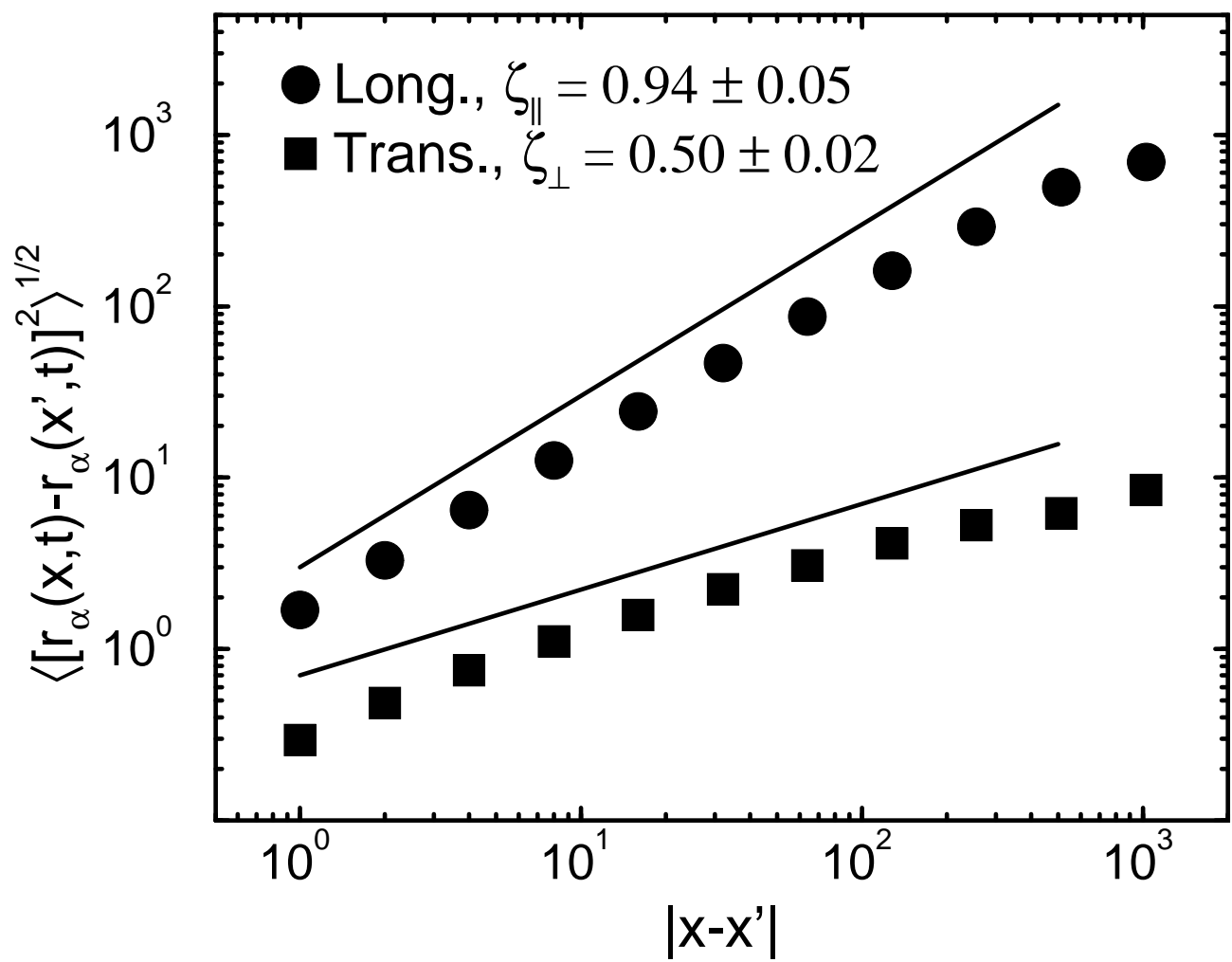
\end{document}